\documentclass[aps,prl,preprint]{revtex4-2}

\usepackage{graphicx}
\usepackage{amsmath}
\usepackage{slashed}
\usepackage{multirow}
\usepackage{subfigure}

\def\etal{{\it et al.}}

\usepackage{xspace}
\usepackage{relsize}
\def\babar{\mbox{\slshape B\kern-0.1em{\smaller A}\kern-0.1em
    B\kern-0.1em{\smaller A\kern-0.2em R}}\xspace}

\def\nn#1#2#3{\hskip -6pt{}^{\phantom{#3}#2}_{\phantom{#2}#3}#1}
\def\degree{^\circ}
\newcommand{\lsim}{\mathrel{\rlap{\lower4pt\hbox{\hskip1pt$\sim$}}
   \raise1pt\hbox{$<$}}}\begin{document}

\title{Flavor-changing Lorentz and CPT violation in muonic atoms}

\author{V.\ Alan Kosteleck\'y,$^1$
W.P.\ McNulty,$^{1}$ 
E.\ Passemar,$^{1,2}$ 
and N.\ Sherrill$^{3,4}$}

\affiliation{$^1$Physics Department, Indiana University, 
Bloomington, Indiana 47405, USA}
\affiliation{$^2$Departament de F\'isica Te\`orica, IFIC, 
Universitat de Val\`encia, E-46071 Val\`encia, Spain}
\affiliation{$^3$Department of Physics and Astronomy, 
University of Sussex, Brighton BN1 9QH, UK}
\affiliation{$^4$Institute of Theoretical Physics, 
Leibniz Universit{\"a}t Hannover, Appelstr.\ 2, 30167 Hannover, Germany}

\date{January 2025}

\begin{abstract}
Flavor-changing signatures of Lorentz and CPT violation 
involving muon-electron conversions in muonic atoms
are studied using effective field theory.
Constraints on coefficients for Lorentz violation 
at parts in $10^{-12}$~GeV$^{-1}$
for flavor-changing electromagnetic muon decays
and parts in $10^{-13}$~GeV$^{-2}$
for flavor-changing 4-point quark-lepton interactions
are extracted using existing data from the SINDRUM~II experiment 
at the Paul Scherrer Institute.
Estimates are provided for sensitivities attainable 
in the forthcoming experiments 
Mu2e at Fermilab
and COMET at the Japan Proton Accelerator Complex.
\end{abstract}

\let\clearpage\relax
\maketitle
\newpage

Flavor-changing processes in the minimal Standard Model (SM)
arise from the weak interactions,
which at tree level
can convert charged leptons to neutrinos 
or mix quark flavors.
The existence of flavor-changing physics beyond the SM 
is revealed by neutrino oscillations~\cite{rd68,sk98},
which establish lepton-number violation.
This implies that flavor-changing conversions between charged leptons 
must also occur in nature,
but only via suppressed one-loop processes
with branching ratios $\lesssim10^{-54}$~\cite{smdecays}.
Experimental searches for charged-lepton flavor-changing (CLFC) processes
therefore offer particularly clean probes 
of additional new physics beyond the SM~\cite{cs18,bh94,bc13}.

One prospective type of CLFC physics beyond the SM
is minuscule violations of Lorentz invariance
and its associated CPT symmetry,
which can emerge at low energies
from an underlying theory unifying gravity and quantum physics
such as strings~\cite{ksp}.
Numerous experimental searches for flavor-conserving Lorentz violation (LV)
have been performed~\cite{tables},
but studies of flavor-changing LV have largely been relegated
to flavor mixing of species during propagation.
However,
the general model-independent framework for LV 
based on effective field theory~\cite{ck97,ak04,kl19,review}
naturally incorporates terms describing CLFC interactions,
of which only a subset contributing to muon decays has been explored 
to date~\cite{ck97,cg99,yi03,gh20,cks20,hp11,nowt16,lmt16,msf17,gmn20,kps22}.
Here,
we use existing data 
to revisit bounds from electromagnetic decays
and to obtain first constraints on
CLFC 4-point quark-lepton interactions. 

A golden channel for experimental studies of CLFC processes,
which currently is the subject of a worldwide experimental effort~\cite{ap23},
is the coherent conversion $\mu + N \to e + N$
of a muon $\mu$ into an electron $e$
in the presence of a nucleus $N$.
The experimental procedure entails the capture of a muon 
in the ground state of a target atom,
followed by its direct conversion into an electron
with a characteristic monoenergetic spectrum.
In model-independent effective field theory for LV searches,
the leading contributions to this channel 
appear already at tree level instead of suppressed loop processes.
No observable CLFC effects in muon decay 
involve LV terms of mass dimension $d=3$ or $4$~\cite{kps22},
so the dominant contributions arise 
via electromagnetic interactions with $d=5$
and 4-point quark-lepton interactions with $d=6$
rather than from standard LV propagator terms~\cite{muon34}.
Since the 4-point LV quark-lepton interactions 
involve the quarks in the nucleus,
the corresponding CLFC effects
are uniquely accessed in the conversion $\mu + N \to e + N$.
No experimental constraints on these effects have been reported to date.
Note that terms with $d=5$ or $6$ appear naturally in certain models 
such as noncommutative quantum field theories~\cite{chklo01}.
Their magnitudes are theoretically undetermined
but are expected either to be suppressed
by the scale of unification or by countershading effects~\cite{kt09}.

In the experiments of interest here,
a proton beam hits a production target,
generating pions and other hadrons that decay to muons and other products. 
The muons with negative charge are collected 
and directed as a secondary beam onto a stopping target,
where some are captured to form muonic atoms.
The muonic atoms quickly decay to the 1S ground state.
The signal for a neutrinoless muon-electron conversion 
is a monoenergetic electron ejected from a muonic atom with energy
$E_e^{\rm conv} = m_\mu - E_{\rm bind} - E_{\rm recoil}$,
where $m_\mu$
is the muon mass, 
$E_{\rm bind}$
is the muon binding energy,
and $E_{\rm recoil}$ 
is the nuclear recoil energy.
This process can be characterized experimentally
by the dimensionless ratio 
$R_{\mu e}$
of the conversion rate $\omega_{\rm conv}$
to the capture rate $\omega_{\rm capt}$.
The dominant contribution to $\omega_{\rm conv}$
is the coherent conversion
$\mu^- + \nn NAZ \rightarrow e^- + \nn NAZ$,
where the nucleus $\nn NAZ$ of the muonic atom 
has atomic number $A$ and charge $Z$.
The ratio $R_{\mu e}$ can then be written as
\begin{equation}
R_{\mu e} = 
\frac{\omega_{\rm conv}}{\omega_{\rm capt}} =
\frac{\Gamma[\mu^- + \nn NAZ \rightarrow e^- + \nn NAZ]}
{\Gamma[(\mu^- + \nn NAZ \rightarrow  \nu_\mu + \nn NA{Z-1})]}.
\label{R}
\end{equation}

To date,
the experiment SINDRUM~II~\cite{sindrum}
at the Paul Scherrer Institute 
has achieved the most stringent constraint,
$R_{\mu e} < 7\times 10^{-13}$ at the $90\%$ confidence level (CL),
using~$\nn{\rm Au}{197}{79}$ as the target. 
Upcoming searches using an~$\nn{\rm Al}{27}{13}$ target
include 
the Muon-to-Electron Conversion (Mu2e) experiment 
at Fermilab~\cite{mu2e},
which anticipates achieving $R_{\mu e} \simeq 6.2\times 10^{-16}$,
and the Coherent Muon to Electron Transition (COMET) 
experiment~\cite{comet}
at the Japan Proton Accelerator Complex (J-PARC),
which expects to reach $R_{\mu e} \simeq 7\times 10^{-15}$.
The forthcoming DeeMe experiment at J-PARC~\cite{deeme},
using graphite and with a different configuration
combining the production and stopping target,
expects sensitivity at the $10^{-14}$ level.
Planned searches for the more distant future include
phase two of Mu2e at Fermilab~\cite{mu2e2},
and phase two of COMET~\cite{comet2} 
and other experiments~\cite{prismprime} at J-PARC,
which are expected to attain $R_{\mu e}$ values of order $10^{-18}$.

Here,
we consider the experiments SINDRUM~II, Mu2e, and COMET, 
which are performed with muonic gold or aluminum atoms.
For$\nn{\rm Au}{197}{79}$ the capture rate is
$\omega_{\rm capt} \simeq 13.07$~MHz,
while for$\nn{\rm Al}{27}{13}$ the capture rate is
$\omega_{\rm capt} \simeq 0.7054$~MHz~\cite{ts87}.  
The capture rate typically increases with $Z$ 
due to a greater overlap between the muon and nucleus wavefunctions, 
but separating the leptonic and nuclear physics 
becomes more challenging for heavier nuclei.
The conversion rate $\omega_{\rm conv} = \overline{|{\cal M}|^2}$ 
is the appropriate spin-averaged squared modulus 
of the transition amplitude ${\cal M}$ for the decay.
We focus here on the coherent contribution
where the nucleus remains in the ground state,
which is experimentally the cleanest channel 
and is enhanced relative to incoherent conversion~\cite{cohconv}.
It has recently been shown that additional sensitivities to CLFC physics
can be accessed from studies of the near-endpoint spectrum~\cite{hr24},
and a corresponding analysis in the LV context
would be a worthwhile topic for future investigation. 
Neglecting corrections below parts in $10^{3}$,
the ground-state nuclear Coulomb field $\rho(r)$ can be taken as
approximately static and spherically symmetric~\cite{nuclsph},
and the nucleus can be treated as infinitely massive
without nuclear recoil and reduced-mass corrections.
The electron energy is then 
$E_{e}^{\rm conv} \approx m_\mu - E_{\rm bind}$ 
where $E_{\rm bind} \ll m_\mu$.

The leptonic wavefunctions can be written in the form 
\begin{equation}
\psi_{\kappa,s} (r, \theta, \phi) = 
{g(r) \chi_{\kappa,s} (\theta,\phi) 
\choose if(r) \chi_{-\kappa,s}(\theta,\phi)},
\end{equation}
where $\kappa = \mp(J+1/2)$ 
and the eigenspinors $\chi_{\kappa, s}$ satisfy 
$(\boldsymbol{\sigma} \cdot \boldsymbol{l}+I)\chi_{\kappa,s}
=-\kappa\chi_{\kappa,s}$ 
and $J_z\chi_{\kappa,s}=s\chi_{\kappa,s}$, 
with normalization 
$\int \,d\Omega~\chi_{\kappa^\prime,s^\prime}
^\dagger\chi_{\kappa,s}
=\delta_{\mu^\prime\mu}\delta_{\kappa^\prime\kappa}$. 
Note that the muon wavefunction in the 1S orbital has $\kappa = -1$,
and also that
that $g^+ = if^-$ and $f^+ = -ig^-$ for a massless electron. 
The radial wavefunctions are obtained by numerically 
solving the Dirac equation 
\begin{equation}
\frac{d}{dr}
\begin{pmatrix} 
u_1 \\ u_2
\end{pmatrix} = 
\begin{pmatrix}
-\kappa/r & W-V(r)+m\\
-(W-V(r)-m) & \kappa/r
\end{pmatrix}
\begin{pmatrix} u_1 \\ u_2\end{pmatrix}
\end{equation}
for $u_1(r)=rg(r)$ and $u_2(r)=rf(r)$.
Here, 
$W$ is the lepton energy
and $V(r)$ is the nuclear electric potential.  
The total muon wavefunction is normalized to unity
and so the muon field has mass dimension $3/2$,
while the electron wavefunction is normalized as a plane wave
and hence the electron field has mass dimension $1$.
The potential $V(r)$ is obtained 
from $\rho(r)$,
which is normalized as $\int_{0}^{\infty} 4\pi\rho(r)r^2 \,dr = Z$
for a nucleus of atomic number $Z$.
The functional form of $\rho(r)$ is taken as 
a two-parameter Fermi model for$\nn{\rm Au}{197}{79}$
and as a Fourier-Bessel expansion for$\nn{\rm Al}{27}{13}$~\cite{vjv87,kko02}.

When the muon-electron conversion is mediated
via LV electromagnetic interactions with $d=5$,
the conversion rate takes the form 
\begin{equation}
\omega_{\rm conv}
= \tfrac{1}{2} \hskip -4pt 
\sum_{s=\pm\frac{1}{2}} 
\sum_{\kappa = \pm 1} 
\sum_{s^\prime=\pm\frac{1}{2}} 
\left| \int d^3x 
F_{\alpha\beta} 
\overline{\psi}{}_{\kappa,s'}^{(e)} 
\mathcal{O}^{\alpha\beta} \psi_{s}^{(\mu)} 
\right|^2.
\label{d5rate}
\end{equation}
Here,
$F_{\alpha\beta}$ is the electromagnetic field strength tensor, 
$\psi_{\kappa,s'}^{(e)}$ 
is a continuum electron wavefunction of energy $W$,
and 
$\psi_{s}^{(\mu)}$ 
is the bound-state muon wavefunction in the 1S state.
The sums implement the effects of the spins and angular momenta,
with the factor of one half arising in averaging the muon spin.
The matrix ${\cal O}^{\alpha\beta}$
contains the $d=5$ coefficients for LV
and appropriate Dirac matrices,
which form the set~\cite{kl19,kps22}
$\{(m_{F}^{(5)})^{\alpha\beta}_{\mu e} $,
$i(m_{5F}^{(5)})^{\alpha\beta}_{\mu e}\gamma_5 $,
$(a_{F}^{(5)})^{\lambda\alpha\beta}_{\mu e}\gamma_\lambda $,
$(b_{F}^{(5)})^{\lambda\alpha\beta}_{\mu e}\gamma_5\gamma_\lambda $,
$\tfrac{1}{2}
(H_{F}^{(5)})^{\kappa\lambda\alpha\beta}_{\mu e}\sigma_{\kappa\lambda} \}$.
These modify interactions rather than propagators,
so standard perturbative quantization methods hold~\cite{kl01}
and the conversion rate \eqref{d5rate}
can be derived using tree-level Feynman rules.
Also, 
in contrast to conventional Lorentz-invariant treatments,
neutrino flavor-changing LV effects~\cite{km12}
can be neglected here because no loop diagrams are involved.

In an inertial frame in the vicinity of the Earth,
the cartesian coefficients for LV 
appearing in ${\cal O}^{\alpha\beta}$
can be taken to be independent of time and location~\cite{ak04}.
The coefficients carry spacetime indices
and so can change under observer Lorentz transformations,
which implies that experimental measurements of their values
must be provided in a specified inertial frame.
The standard choice in the literature is the Sun-centered frame (SCF)
with right-handed cartesian coordinates $(T,X,Y,Z)$,
where the time origin $T=0$ is defined as the 2000 vernal equinox,
the $X$ axis is chosen to point from the Earth to the Sun at the time zero,
and the $Z$ axis parallels the rotation axis of the Earth~\cite{sunframe}.  
We therefore must determine the transformation from the SCF 
to the detector frame (DF). 
For this purpose,
it is convenient to fix also a standard Earth-based laboratory frame (LF),
with right-handed cartesian coordinates $(x,y,z)$
having $x$ axis pointing to local south and $y$ axis to local east.

Any laboratory frame distant from the poles is noninertial
due to the rotation of the Earth, 
which has sidereal frequency
${\omega_\oplus}\simeq 2\pi /(23\,{\rm h}~56\,\min)$.
The boost of the Earth in the SCF is small,
so at leading order the transformation from the SCF to the LF 
is a rotation $\mathcal{R}(\chi, \omega_\oplus T_\oplus)$
that depends on the laboratory colatitude~$\chi$ 
and is harmonic in the laboratory sidereal time
$T_\oplus \equiv T - T_0$.
The time $T_\oplus$ in the LF 
is shifted by an experiment-dependent amount $T_0$
relative to the time $T$ in the SCF,
arising from the laboratory longitude $\lambda$ 
and other effects~\cite{offset}.
The explicit form of the rotation $\mathcal{R}(\chi, \omega_\oplus T_\oplus)$ 
is given as Eq.~(7) of Ref.~\cite{kps22},
and it can be used to transform coefficient components
from the SCF to the LF.
The DF typically differs from the standard LF,
so we must perform an additional transformation 
$\mathcal{R_{\text{detector}}}(\psi)$
given in Eq.~(6) of Ref.~\cite{kps22}
that involves the angle $\psi$ of the laboratory $z$-axis
measured north of east along the direction of the beamline.
The quantities $\chi$, $\lambda$, $T_0$, and $\psi$
for the experiments considered here are listed 
in Table~\ref{exptprops}.

\begin{table*}
\centering
\setlength{\tabcolsep}{5pt}
\begin{tabular}{c | c c c c c c c c c}
\hline\hline
Experiment $	$&$	\chi	$&$	\lambda	$&$	T_0	$&$	\psi	$&$	c	$&$	\widetilde{I}_1	$&$	\widetilde{I}_2	$&$	\widetilde{I}_3	$&$	\widetilde{I}_4	$\\	\hline
SINDRUM II $	$&$	42.5\degree	$&$	8.2\degree	$&$	3.86\;\rm{h} 	$&$	242\degree	$&$	0.44	$&$	-0.053	$&$	0.070	$&$	0.040	$&$	0.18	$\\
COMET $	$&$	53.6\degree	$&$	140.6\degree	$&$	-4.94\;\rm{h} 	$&$	188\degree	$&$	0.34	$&$	-0.012	$&$	0.012	$&$	7.6\times 10^{-4}	$&$	0.041	$\\
Mu2e $	$&$	48.2\degree	$&$	-88.2\degree	$&$	10.27\;\rm{h} 	$&$	122\degree	$&$	0.50	$&$	-0.012	$&$	0.012	$&$	7.6\times 10^{-4}	$&$	0.041	$\\
\hline\hline 
\end{tabular}
\caption{Detector geometric factors, parameters, and radial integrals.}
\label{exptprops}
\end{table*}

The above considerations show that
the net transformation $\mathcal{R_{\text{total}}}$ 
from the SCF to the DF is given by the combination 
\begin{equation}
\mathcal{R_{\text{total}}} = 
\mathcal{R_{\text{detector}}}(\psi) 
\mathcal{R}(\chi, \omega_\oplus T_\oplus).
\label{map}
\end{equation}
The components of the coefficients for LV 
expressed in the DF
therefore depend on the detector orientation and location,
and they can oscillate at harmonics of the sidereal frequency~\cite{ak98}.
This reveals that the conversion rate 
and the ratio $R_{\mu e}$ 
have corresponding dependences,
$\omega_{\rm conv} = \omega_{\rm conv} (\psi, \chi, \omega_\oplus T_\oplus)$
and 
$R_{\mu e} = R_{\mu e} (\psi, \chi, \omega_\oplus T_\oplus)$.
An experiment recording data with time stamps
can measure the amplitudes and phases of the harmonics
by binning the data in sidereal time
and hence can detect or constrain
the various components of the coefficients for LV in the SCF. 

The spherical symmetry of the nuclear charge distribution
makes it convenient for some calculations
to adopt DF spherical-polar coordinates $(r, \theta, \phi)$
with $z$ axis directed along the muon beam direction.
Results must then be matched to DF cartesian coordinates
to apply the map \eqref{map} from the SCF to the DF. 
The spherical symmetry also
implies that the nonzero contributions to the rate~\eqref{d5rate}
from the field-strength tensor arise only from the components
$F_{tr} = -F_{rt} = E_r(r)$,
where the index $t$ represents the time $T_\oplus$. 
Like $F_{\alpha\beta}$,
the matrix ${\cal O}^{\alpha\beta}$ is antisymmetric in $\alpha\beta$,
so it suffices to consider the case $\alpha = t$, $\beta = r$.
The matrix component ${\cal O}^{tr}$ is a spatial vector,
so the match between polar and cartesian coordinates in the DF
can be accomplished by 
$\mathcal{O}^{tr} = 
\mathcal{O}^{tx} \sin{\theta}\cos{\phi}
+\mathcal{O}^{ty} \sin{\theta}\sin{\phi}
+\mathcal{O}^{tz} \cos{\theta}$,
as usual.
These additional angular factors must be taken into account
when performing the integration in Eq.~\eqref{d5rate}
to obtain the conversion rate $\omega_{\rm conv}$.

Working in DF spherical-polar coordinates,
the integral \eqref{d5rate}
can be taken over the full azimuthal range $\phi\in[0,2\pi)$.
However,
the angle $\theta$ is limited by the detector acceptance
to a range $|\cos\theta|\lsim c$.
For the experiments considered here,
approximate values of the acceptances $c$ 
are listed in Table~\ref{exptprops}.
For the angular part of the integral~\eqref{d5rate},
the choices $\kappa=\pm1$, $s=\pm 1/2$, and $s'=\pm1/2$ 
yield eight possible cases in principle.
However,
for any given coefficient for LV,
only two nonzero cases occur in practice. 
They depend on the acceptance $c$ for the given experiment,
via two geometrical factors 
$w_{xy} = -c^3/6+c/2$ and $w_z = c^3/3$.
For the radial part two integrals of mass dimension 3/2 occur,
\begin{align}
I_1 =& \int \,dr~r^2 E(r) \left(f_e^-g_\mu^-+g_e^-f_\mu^-\right),
\nonumber\\
I_2 =& \int \,dr~r^2 E(r) \left(g_e^-f_\mu^--f_e^-g_\mu^-\right).
\label{integrals}
\end{align}
These integrals also take unique values for each experiment.
Table \ref{exptprops} displays numerical values of the quantities 
$\widetilde{I}_1 = I_1/m_\mu^{3/2}$ and $\widetilde{I}_2 = I_2/m_\mu^{3/2}$
for the various experiments we analyze here.
Combining these results and converting to cartesian coordinates
gives the desired equations for the transition amplitudes ${\cal M}$
and hence for the conversion rate $\omega_{\rm conv}$
expressed using coefficients for LV in the DF.
Applying the transformation \eqref{map}
then yields $\omega_{\rm conv}$ in terms of coefficients for LV in the SCF.
Explicit forms for these results are presented in the Appendix.

The SINDRUM~II bound 
was obtained using a primary dataset taken 
over a period of 81 days~\cite{sindrum}.
It can therefore be intepreted as a limit 
on the time-averaged conversion rate 
$\overline{\omega}_{\rm conv} = 
\overline{\omega}_{\rm conv} (\psi, \chi)$
for the corresponding $\psi$ and $\chi$ values
given in Table~\ref{exptprops}.
Adopting the standard methodology in the literature~\cite{tables},
we can translate this experimental limit into constraints
on the coefficients for LV in the SCF,
taken one at a time. 
The key expressions involved in this procedure
are presented in the Appendix,
and the ensuing constraints are tabulated in Table~\ref{resultsd5}.
In the table,
each entry is a bound at 90\% CL 
on a coefficient component in the SCF,
where the indices $J$ and $K\neq J$ are $X$ or $Y$.
The table reveals that SINDRUM~II achieved sensitivities
down to a few parts in $10^{12}$ 
to 96 real components of the coefficients for LV at $d=5$.
Note that the constraints involving the coefficients 
$(a_{F}^{(5)})^{\lambda\alpha\beta}_{\mu e}$ and
$(b_{F}^{(5)})^{\lambda\alpha\beta}_{\mu e}$
also represent limits on CPT violation. 
The same methodology can be applied
to the anticipated reaches of the Mu2e and COMET experiments,
yielding the estimated attainable sensitivities
displayed in Table~\ref{resultsd5}.
Together with the enhanced resolving power arising 
via the separation of harmonics in sidereal time,
these estimates suggest improvements of one to two orders of magnitude
over the SINDRUM~II results are feasible within the near future.

\renewcommand\arraystretch{0.8}
\begin{table}
\centering
\setlength{\tabcolsep}{2pt}
\begin{tabular}{c c c c c}
\hline\hline								
Coefficients	&	SINDRUM~II	&	COMET &	Mu2e	\\	
\hline								
$ |(m_{F}^{(5)})^{TJ}_{\mu e}|, |(m_{5F}^{(5)})^{TJ}_{\mu e}|	$&$	< 8	$&$	< 1	$&$	< 0.2	$ \\
$ |(m_{F}^{(5)})^{TZ}_{\mu e}|, |(m_{5F}^{(5)})^{TZ}_{\mu e}|	$&$	< 8	$&$	< 0.9	$&$	< 0.2	$ \\ [4pt]
$ |(a_{F}^{(5)})^{TTJ}_{\mu e}|, |(b_{F}^{(5)})^{TTJ}_{\mu e}|	$&$	< 6	$&$	< 1	$&$	< 0.2	$ \\
$ |(a_{F}^{(5)})^{TTZ}_{\mu e}|, |(b_{F}^{(5)})^{TTZ}_{\mu e}|	$&$	< 6	$&$	< 0.9	$&$	< 0.2	$ \\
$ |(a_{F}^{(5)})^{JTJ}_{\mu e}|, |(b_{F}^{(5)})^{JTJ}_{\mu e}|	$&$	< 6	$&$	< 1	$&$	< 0.2	$ \\
$ |(a_{F}^{(5)})^{JTK}_{\mu e}|, |(b_{F}^{(5)})^{JTK}_{\mu e}|	$&$	< 8	$&$	< 1	$&$	< 0.2	$ \\ 
$ |(a_{F}^{(5)})^{JTZ}_{\mu e}|, |(b_{F}^{(5)})^{JTZ}_{\mu e}|	$&$	< 8	$&$	< 0.9	$&$	< 0.2	$ \\
$ |(a_{F}^{(5)})^{ZTJ}_{\mu e}|, |(b_{F}^{(5)})^{ZTJ}_{\mu e}|	$&$	< 8	$&$	< 1	$&$	< 0.2	$ \\
$ |(a_{F}^{(5)})^{ZTZ}_{\mu e}|, |(b_{F}^{(5)})^{ZTZ}_{\mu e}|	$&$	< 7	$&$	< 0.9	$&$	< 0.2	$ \\ [4pt]
$ |(H_{F}^{(5)})^{TJTJ}_{\mu e}|, |(H_{F}^{(5)})^{JZTK}_{\mu e}|	$&$	< 7	$&$	< 1	$&$	< 0.2	$ \\
$ |(H_{F}^{(5)})^{TJTK}_{\mu e}|, |(H_{F}^{(5)})^{JZTJ}_{\mu e}|	$&$	< 6	$&$	< 1	$&$	< 0.2	$ \\
$ |(H_{F}^{(5)})^{TJTZ}_{\mu e}|, |(H_{F}^{(5)})^{JZTZ}_{\mu e}|	$&$	< 6	$&$	< 0.9	$&$	< 0.2	$ \\ 
$ |(H_{F}^{(5)})^{TZTJ}_{\mu e}|, |(H_{F}^{(5)})^{XYTJ}_{\mu e}|	$&$	< 6	$&$	< 1	$&$	< 0.2	$ \\
$ |(H_{F}^{(5)})^{TZTZ}_{\mu e}|, |(H_{F}^{(5)})^{XYTZ}_{\mu e}|	$&$	< 7	$&$	< 0.9	$&$	< 0.2	$ \\[4pt]
\hline\hline 
\end{tabular}
\caption{Constraints on $d=5$ coefficients for LV 
in units of $10^{-12}$ GeV$^{-1}$.
The results for SINDRUM~II are constraints at 90\% CL. 
The results for Mu2e and COMET are projected constraints 
based on expected rate sensitivities.}
\label{resultsd5}
\end{table}

\renewcommand\arraystretch{0.8}
\begin{table}
\centering
\setlength{\tabcolsep}{3pt}
\begin{tabular}{c c c c c}
\hline\hline								
Coefficients	&	SINDRUM~II	&	COMET &	Mu2e	\\	
\hline								
$ |(k_{SV}^{(6)})^T_{uue\mu}|, |(k_{SA}^{(6)})^T_{uue\mu}|	$&$	<6	$&$	<1	$&$	<0.2	$ \\
$ |(k_{SV}^{(6)})^J_{uue\mu}|, |(k_{SA}^{(6)})^J_{uue\mu}|	$&$	<7	$&$	<1	$&$	<0.2	$ \\
$ |(k_{SV}^{(6)})^Z_{uue\mu}|, |(k_{SA}^{(6)})^Z_{uue\mu}|	$&$	<7	$&$	<1	$&$	<0.2	$ \\
$ |(k_{SV}^{(6)})^T_{dde\mu}|, |(k_{SA}^{(6)})^T_{dde\mu}|	$&$	<6	$&$	<1	$&$	<0.2	$ \\
$ |(k_{SV}^{(6)})^J_{dde\mu}|, |(k_{SA}^{(6)})^J_{dde\mu}|	$&$	<7	$&$	<1	$&$	<0.2	$ \\
$ |(k_{SV}^{(6)})^Z_{dde\mu}|, |(k_{SA}^{(6)})^Z_{dde\mu}|	$&$	<7	$&$	<1	$&$	<0.2	$ \\ 
$ |(k_{SV}^{(6)})^T_{sse\mu}|, |(k_{SA}^{(6)})^T_{sse\mu}|	$&$	<10	$&$	<2	$&$	<0.4	$ \\
$ |(k_{SV}^{(6)})^J_{sse\mu}|, |(k_{SA}^{(6)})^J_{sse\mu}|	$&$	<15	$&$	<2	$&$	<0.4	$ \\
$ |(k_{SV}^{(6)})^Z_{sse\mu}|, |(k_{SA}^{(6)})^Z_{sse\mu}|	$&$	<15	$&$	<2	$&$	<0.4	$ \\
$ |(k_{VS}^{(6)})^{T}_{uu e\mu}|, |(k_{VP}^{(6)})^{T}_{uu e\mu}|	$&$	<30	$&$	<4	$&$	<0.8	$ \\
$ |(k_{VS}^{(6)})^{T}_{dde\mu}|, |(k_{VP}^{(6)})^{T}_{dde\mu}|	$&$	<30	$&$	<4	$&$	<0.7	$ \\ [4pt]
$|(k_{ST}^{(6)})^{TJ}_{uue\mu}|, |(k_{ST}^{(6)})^{JZ}_{uue\mu}|	$&$	<7	$&$	<1	$&$	<0.2	$ \\ 
$|(k_{ST}^{(6)})^{TZ}_{uue\mu}|, |(k_{ST}^{(6)})^{XY}_{uue\mu}|	$&$	<7	$&$	<1	$&$	<0.2	$ \\
$|(k_{ST}^{(6)})^{TJ}_{dde\mu}|, |(k_{ST}^{(6)})^{JZ}_{dde\mu}|	$&$	<7	$&$	<1	$&$	<0.2	$ \\ 
$|(k_{ST}^{(6)})^{TZ}_{dde\mu}|, |(k_{ST}^{(6)})^{XY}_{dde\mu}|	$&$	<7	$&$	<1	$&$	<0.2	$ \\ 
$|(k_{ST}^{(6)})^{TJ}_{sse\mu}|, |(k_{ST}^{(6)})^{JZ}_{sse\mu}|	$&$	<15	$&$	<2	$&$	<0.4	$ \\ 
$|(k_{ST}^{(6)})^{TZ}_{sse\mu}|, |(k_{ST}^{(6)})^{XY}_{sse\mu}|	$&$	<15	$&$	<2	$&$	<0.4	$ \\ 
$|(k_{VV}^{(6)})^{TT}_{uue\mu}|, |(k_{VA}^{(6)})^{TT}_{uue\mu}|	$&$	<20	$&$	<4	$&$	<0.7	$ \\ 
$|(k_{VV}^{(6)})^{TJ}_{uue\mu}|, |(k_{VA}^{(6)})^{TJ}_{uue\mu}|	$&$	<25	$&$	<4	$&$	<0.7	$ \\ 
$|(k_{VV}^{(6)})^{TZ}_{uue\mu}|, |(k_{VA}^{(6)})^{TZ}_{uue\mu}|	$&$	<25	$&$	<4	$&$	<0.7	$ \\ 
$|(k_{VV}^{(6)})^{TT}_{dde\mu}|, |(k_{VA}^{(6)})^{TT}_{dde\mu}|	$&$	<20	$&$	<4	$&$	<0.7	$ \\ 
$|(k_{VV}^{(6)})^{TJ}_{dde\mu}|, |(k_{VA}^{(6)})^{TJ}_{dde\mu}|	$&$	<20	$&$	<4	$&$	<0.7	$ \\ 
$|(k_{VV}^{(6)})^{TZ}_{dde\mu}|, |(k_{VA}^{(6)})^{TZ}_{dde\mu}|	$&$	<20	$&$	<4	$&$	<0.7	$ \\ [4pt]
$|(k_{VT}^{(6)})^{TTJ}_{uue\mu}|, |(k_{VT}^{(6)})^{TJZ}_{uue\mu}|	$&$	<25	$&$	<4	$&$	<0.7	$ \\ 
$|(k_{VT}^{(6)})^{TTZ}_{uue\mu}|, |(k_{VT}^{(6)})^{TXY}_{uue\mu}|	$&$	<25	$&$	<4	$&$	<0.7	$ \\ 
$|(k_{VT}^{(6)})^{TTJ}_{dde\mu}|, |(k_{VT}^{(6)})^{TJZ}_{dde\mu}|	$&$	<20	$&$	<4	$&$	<0.7	$ \\ 
$|(k_{VT}^{(6)})^{TTZ}_{dde\mu}|, |(k_{VT}^{(6)})^{TXY}_{dde\mu}|	$&$	<20	$&$	<4	$&$	<0.7	$ \\ [4pt]
\hline\hline
\end{tabular}
\caption{Constraints on $d=6$ coefficients for LV 
in units of $10^{-13}$ GeV$^{-2}$.
The results for SINDRUM~II are constraints at 90\% CL. 
The results for Mu2e and COMET are projected constraints 
based on expected rate sensitivities.}
\label{resultsd6}
\end{table}

If instead the muon-electron conversion is mediated
through 4-point interactions coupling the nuclear quarks to the leptons
with $d=6$,
the conversion rate is 
\begin{equation}
\omega_{\rm conv} 
= \tfrac{1}{2} \hskip -4pt 
\sum_{s,s',\kappa}  
\left| \int d^3x 
\left(
\alpha
\overline{\psi}{}_{\kappa,s'}^{(e)} 
\mathcal{K} \psi_{s}^{(\mu)}
+
\beta
\overline{\psi}{}_{\kappa,s'}^{(e)} 
\mathcal{K}_0 \psi_{s}^{(\mu)}
\right)
\right|^2,
\label{d6rate}
\end{equation}
where $\alpha$ and $\beta$ are conventional nuclear matrix elements
given by
$\alpha = \langle N| \overline{\psi}{}^{(q)} \psi^{(q)} |N\rangle$
with $q$ summed over the quark flavors $q = u,d,s$
and 
$\beta = \langle N| \overline{\psi}{}^{(q)} \gamma_0 \psi^{(q)}|N \rangle$
with $q$ summed over $q = u,d$.
Other possible nuclear matrix elements vanish 
in coherent conversion~\cite{kko02}.
The operators ${\cal K}$ and ${\cal K}_0$ 
incorporate the $d=6$ coefficients for LV 
and appropriate Dirac matrices~\cite{kl19},
with ${\cal K}$ drawn from the set 
$\{(k_{SV}^{(6)})^{\lambda}_{qqe\mu}\gamma_\lambda$,
$(k_{SA}^{(6)})^{\lambda}_{qqe\mu}\gamma_5\gamma_\lambda$,
$\tfrac{1}{2}
(k_{ST}^{(6)})^{\kappa\lambda}_{qqe\mu}\sigma_{\kappa\lambda}\}$
for $q = u,d,s$
and 
${\cal K}_0$ 
from 
$\{(k_{VS}^{(6)})^{t}_{qqe\mu}$,
$i(k_{VP}^{(6)})^{t}_{qqe\mu}\gamma_5$,
$(k_{VV}^{(6)})^{t\lambda}_{qqe\mu}\gamma_\lambda$,
$(k_{VA}^{(6)})^{t\lambda}_{qqe\mu}\gamma_5\gamma_\lambda$,
$\tfrac{1}{2}
(k_{VT}^{(6)})^{t\kappa\lambda}_{qqe\mu}\sigma_{\kappa\lambda}\}$
for $q = u,d$.
These arise from 4-point quark-lepton interactions 
in the underlying effective field theory,
for which the standard perturbative quantization
yields the expression \eqref{d6rate} 
arising from vertex contributions to the Feynman rules.

In DF spherical-polar coordinates,
the angular part of the integral \eqref{d6rate}
involves eight possible cases
$\kappa=\pm1$, $s=\pm 1/2$, and $s'=\pm1/2$.
Their values depend on the acceptance $c$ for the given experiment.
The radial part involves two integrals of mass dimension 5/2,
\begin{equation}
I_3 = \int \,dr~r^2 \rho^{(p)}(r) f_e^-f_\mu^-,
\quad
I_4 = \int \,dr~r^2 \rho^{(p)}(r) g_e^-g_\mu^-.
\label{integrals2}
\end{equation}
For the experiments considered here,
numerical values of 
$\widetilde{I}_3 = I_3/m_\mu^{5/2}$ and $\widetilde{I}_4 = I_4/m_\mu^{5/2}$
are given in Table~\ref{exptprops}.
These equations permit evaluation of the transition amplitudes ${\cal M}$
and hence of $\omega_{\rm conv}$ 
expressed using coefficients for LV in the DF.
The result for $\omega_{\rm conv}$ 
in terms of coefficients for LV in the SCF 
can then be obtained by applying the transformation~\eqref{map}.
Some details of this procedure are provided in the Appendix.

Interpreting the limit from SINDRUM~II as a bound
on the time-averaged conversion rate $\overline{\omega}_{\rm conv}$,
we can establish first constraints in the SCF 
on each component of coefficients for $d=6$ LV taken in turn.
These constraints are presented in Table~\ref{resultsd6},
where $J$ is $X$ or $Y$.
The results show that SINDRUM~II attained sensitivities
down to parts in $10^{13}$ 
to CLFC effects from 148 real components of coefficients for LV
arising from 4-point quark-lepton interactions.
Note that bounds on coefficients with one or three spacetime indices
are also constraints on CPT violation. 
Table~\ref{resultsd6} also displays projected sensitivities 
in the Mu2e and COMET experiments.
Allowing also for analyses of the harmonic sidereal-time variations,
these results suggest that impressive enhancements in sensitivities
and a corresponding significant discovery potential
lie within future reach.

\medskip
 
This work is supported in part by 
the U.S.\ Department of Energy 
under grants {DE}-SC0010120 and {DE}-AC05-06OR23177,
by the U.S.\ National Science Foundation 
under grant PHY-2013184,
by the Generalitat Valenciana (Spain) 
through the plan GenT program CIDEGENT/2021/037,
by the Spanish Government Agencia Estatal de Investigaci\'on 
MCIN/AEI/10.13039/501100011033
under grants PID2020–114473GB-I00 and PID2023-146220NB-I00,
by the Agencia Estatal de Investigaci\'on MCIU/AEI (Spain)
under grant IFIC Centro de Excelencia Severo Ochoa CEX2023-001292-S,
by the U.K.\ Science and Technology Facilities Council
under grants ST/T006048/1 and ST/Y004418/1,
by the Deutsche Forschungsgemeinschaft 
under the Heinz Maier Leibnitz Prize BeyondSM HML-537662082,
and by the Indiana University Center for Spacetime Symmetries.

\medskip

\newpage

{\bf Appendix: conversion rates.}
The conversion rate via $d=5$ LV electromagnetic interactions
is given by Eq.~\eqref{d5rate}
and can be evaluated in the DF using spherical polar coordinates.
Combining the geometrical factors 
$w_{xy}$ and $w_z$ 
with the radial integrals~\eqref{integrals}
for the allowed values of the 
muon spin $s=\pm 1/2$, 
electron spin $s'=\pm1/2$,
and electron angular momentum $\kappa=\pm1$
yields expressions for the transition amplitudes 
${\cal{M}}^{\kappa}_{s',s}$
in cartesian coordinates in the DF,
\begin{align}
{\cal{M}}^{+1}_{\pm\frac12,\pm\frac12}
&=
\mp i w_{z}(m_F^{(5)})_{\mu e}^{tz}I_1
\pm iw_{z}(a_F^{(5)})_{\mu e}^{ttz}I_2
\nonumber\\
&
\pm w_{xy}((b_F^{(5)})_{\mu e}^{xty}-(b_F^{(5)})_{\mu e}^{ytx})I_1
\nonumber\\
&
-i[w_{xy}((b_F^{(5)})_{\mu e}^{xtx}+(b_F^{(5)})_{\mu e}^{yty})+w_{z}(b_F^{(5)})_{\mu e}^{ztz}]I_2
\nonumber\\
&
\pm w_{xy}((H_F^{(5)})_{\mu e}^{xztx}+(H_F^{(5)})_{\mu e}^{yzty})I_2
\nonumber\\
&+
i[w_{xy}((H_F^{(5)})_{\mu e}^{xzty}-(H_F^{(5)})_{\mu e}^{yztx})
\nonumber\\
&\quad
-w_{z}(H_F^{(5)})_{\mu e}^{xytz}]I_1,
\end{align}
\begin{align}
{\cal{M}}^{+1}_{\mp\frac12,\pm\frac12}
&=
-iw_{xy}((m_F^{(5)})_{\mu e}^{tx} \pm i(m_F^{(5)})_{\mu e}^{ty})I_1
\nonumber\\
&+
iw_{xy}((a_F^{(5)})_{\mu e}^{ttx}  \pm i (a_F^{(5)})_{\mu e}^{tty})I_2
\nonumber\\
&+
[(w_{z}(b_F^{(5)})_{\mu e}^{ytz} - w_{xy}(b_F^{(5)})_{\mu e}^{zty})
\nonumber\\
&\quad
\pm i(w_{z}(b_F^{(5)})_{\mu e}^{xtz}-w_{xy}(b_F^{(5)})_{\mu e}^{ztx})]I_1
\nonumber\\
&
-[(w_{xy}(H_F^{(5)})_{\mu e}^{xyty}+w_{z}(H_F^{(5)})_{\mu e}^{xztz})
\nonumber\\
&\quad
\mp i(w_{xy}(H_F^{(5)})_{\mu e}^{xytx} - w_{z}(H_F^{(5)})_{\mu e}^{yztz})]I_2,
\end{align}
\begin{align}
{\cal{M}}^{-1}_{\pm\frac12,\pm\frac12}
&=
\pm w_{z}(m_{5F}^{(5)})_{\mu e}^{tz}I_1
\nonumber\\
&
\pm w_{xy}((a_F^{(5)})_{\mu e}^{xty} - (a_F^{(5)})_{\mu e}^{ytx})I_1
\nonumber\\
&
-i[w_{xy}((a_F^{(5)})_{\mu e}^{xtx}+(a_F^{(5)})_{\mu e}^{yty}) 
\nonumber\\
&\quad
+w_{z}(a_F^{(5)})_{\mu e}^{ztz}]I_2
\pm iw_{z} (b_F^{(5)})_{\mu e}^{ttz}I_2
\nonumber\\
&
\pm iw_{xy}((H_F^{(5)})_{\mu e}^{txty}-(H_F^{(5)})_{\mu e}^{tytx})I_2
\nonumber\\
&
+ [w_{xy}((H_F^{(5)})_{\mu e}^{txtx}+(H_F^{(5)})_{\mu e}^{tyty})
\nonumber\\
&\quad
+w_{z}(H_F^{(5)})_{\mu e}^{tztz}]I_1,
\end{align}
\begin{align}
{\cal{M}}^{-1}_{\mp\frac12,\pm\frac12}
&=
w_{xy}((m_{5F}^{(5)})_{\mu e}^{tx} \pm i (m_{5F}^{(5)})_{\mu e}^{ty})I_1
\nonumber\\
&+
[(w_{z}(a_F^{(5)})_{\mu e}^{ytz} - w_{xy}(a_F^{(5)})_{\mu e}^{zty})
\nonumber\\
&\quad
\pm i(w_{z}(a_F^{(5)})_{\mu e}^{xtz}-w_{xy}(a_F^{(5)})_{\mu e}^{ztx})]I_1
\nonumber\\
&+
iw_{xy}((b_F^{(5)})_{\mu e}^{ttx} \pm i (b_F^{(5)})_{\mu e}^{tty})I_2
\nonumber\\
&
\pm [(w_{z}(H_F^{(5)})_{\mu e}^{txtz}-w_{xy}(H_F^{(5)})_{\mu e}^{tztx})
\nonumber\\
&\quad
+i(w_{z}(H_F^{(5)})_{\mu e}^{tytz}-w_{xy}(H_F^{(5)})_{\mu e}^{tzty})]I_2.
\end{align}
The conversion rate for any desired combination of coefficients in the DF 
can be obtained from these equations as 
$\omega_{\rm conv} = \overline{|{\cal M}|^2}$.
 
To determine $\omega_{\rm conv}$ 
in terms of coefficients for LV in the SCF,
where the coefficients are constants
and useful constraints can thus be established,
we apply the map \eqref{map}.
This produces a lengthy expression
containing harmonics of $\omega_\oplus T_\oplus$.
For applications to datasets taken over extended time intevals,
time averaging can be performed,
which leaves only the time-independent contribution 
$\overline{\omega}_{\rm conv}$.
An experimental measurement of $\overline{\omega}_{\rm conv}$
can thereby be translated into a limit 
on the components of the coefficients for LV
taken one at a time.
The constraints from SINDRUM~II
and the estimated sensitivities of Mu2e and COMET
resulting from this procedure
are presented in Table~\ref{resultsd5}.

As an illustration,
consider the coefficient $(m_{F}^{(5)})^{\alpha\beta}_{\mu e}$.
In the DF,
we find
\begin{align}
{\omega}_{\rm conv} = 
\left(
w_{xy}^2
\textstyle{\sum}_j |(m_F^{(5)})_{\mu e}^{tj}|^2 
+w_z^2
|(m_F^{(5)})_{\mu e}^{tz}|^2 
\right) I_1^2 ,
\label{sindrum-rate}
\end{align}
where $j=x,y$.
Converting to the SCF and taking the time average yields
$\overline{\omega}_{\rm conv} = \zeta_{TJ}^{(0)} 
|(m_F^{(5)})_{\mu e}^{TJ}|^2$,
where $J=X,Y$ 
and $\zeta_{TJ}^{(0)}$ is an experiment-dependent quantity 
that depends on the geometric factors in Table~\ref{exptprops}.
For SINDRUM~II,
for example,
we find $\zeta_{TJ}^{(0)}\simeq 9.5\times 10^{-8}$ GeV$^3$.
This result provides the constraints
on the coefficient components $(m_F^{(5)})_{\mu e}^{TJ}$
listed in the first row of Table~\ref{resultsd5}.

For completeness,
we provide here the terms in $\omega_{\rm conv}$ 
involving the squared moduli of each coefficient for LV
expressed in the SCF,
\begin{align}
\omega_{\rm conv}
&\supset 
( \zeta_{TJ}^{(0)} 
+ \zeta_{TJ}^{(2c)} c_{2\omega_\oplus T_\oplus} 
+ \zeta_{TJ}^{(2s)} s_{2\omega_\oplus T_\oplus} )
\nonumber\\
&\qquad\times
(|(m_F^{(5)})_{\mu e}^{TJ}|^2 
+ |(m_{5F}^{(5)})_{\mu e}^{TJ}|^2 ) 
\nonumber\\ &
+( \zeta_{\alpha TJ}^{(0)} 
+ \zeta_{\alpha TJ}^{(2c)} c_{2\omega_\oplus T_\oplus} 
+ \zeta_{\alpha TJ}^{(2s)} s_{2\omega_\oplus T_\oplus} 
\nonumber\\
&\qquad
+ \zeta_{\alpha TJ}^{(4c)} c_{4\omega_\oplus T_\oplus} 
+ \zeta_{\alpha TJ}^{(4s)} s_{4\omega_\oplus T_\oplus} )
\nonumber\\
&\qquad\qquad\times
( |(a_F^{(5)})_{\mu e}^{\alpha TJ}|^2 
+ |(b_F^{(5)})_{\mu e}^{\alpha TJ}|^2 )
\nonumber\\ &   
+( \zeta_{\alpha \beta TJ}^{(0)} 
+ \zeta_{\alpha \beta TJ}^{(2c)} c_{2\omega_\oplus T_\oplus} 
+ \zeta_{\alpha \beta TJ}^{(2s)} s_{2\omega_\oplus T_\oplus} 
\nonumber\\
&\qquad
+ \zeta_{\alpha \beta TJ}^{(4c)} c_{4\omega_\oplus T_\oplus} 
+ \zeta_{\alpha \beta TJ}^{(4s)} s_{4\omega_\oplus T_\oplus} )
|(H_F^{(5)})_{\mu e}^{\alpha \beta TJ}|^2,
\label{omconv5}
\end{align}
where $c_{n\omega_\oplus T_\oplus} = \cos{n\omega_\oplus T_\oplus}$
and $s_{n\omega_\oplus T_\oplus} = \sin{n\omega_\oplus T_\oplus}$
for $n=2,4$.
Terms in the full expression that are omitted above
involve products of two different coefficients for LV,
which are irrelevant for constraints obtained
with one coefficient component taken nonzero at a time.

If instead the conversion is mediated 
by $d=6$ LV operators governing 4-point quark-lepton interactions,
the conversion rate is given by Eq.~\eqref{d6rate}.
The values of the conventional nuclear matrix elements
$\alpha$ and $\beta$ can be expressed
in terms of the nuclear proton density $\rho^{(p)}$,
the neutron density $\rho^{(n)}$,
$A$, $Z$, 
and numerical parameters $G_S^{(q,p)}$, $G_S^{(q,n)}$~\cite{kko02}.
The latter take the values
$G_S^{(q,p)} = \{5.1,4.3,2.5\}$ 
and $G_S^{(q,n)} = \{4.3,5.1,2.5\}$ 
for $q = \{u,d,s\}$, 
respectively~\cite{kks01}.
We take
$\rho^{(p)}$ to be normalized to $Z$ 
and $\rho^{(n)}$ to follow $\rho^{(p)}$
but to be normalized to $A-Z$,
and we define 
$\widetilde{G}_S^q \equiv [ZG_S^{(q,p)}+(A-Z)G_S^{(q,n)}]/Z$, 
$\widetilde{G}_V^u \equiv (Z+A)/Z$, 
and $\widetilde{G}_V^d \equiv (2A-Z)/Z$.
Combining these results with geometric factors
and the integrals~\eqref{integrals2}
permits expressing the transition amplitude $\cal M$
in terms of the combinations
$I_{S\pm}^q = \pm c \widetilde{G}_S^q (I_3\pm I_4)$,
$I_{S1}^q = c \widetilde{G}_S^q 
(\tfrac{1}{3} c^2 I_3+I_4)$,
and $I_{S2}^q = c \widetilde{G}_S^q
[(1-\tfrac{2}{3} c^2)I_3+I_4]$
for $q = u,d,s$,
along with
$I_{V\pm}^{q} = \pm c \widetilde{G}_V^q (I_3\pm I_4)$,
$I_{V1}^{q} = c \widetilde{G}_V^q \left(\tfrac{1}{3} c^2 I_3+I_4\right)$,
and $I_{V2}^{q} = c \widetilde{G}_V^q [(1-\tfrac{2}{3} c^2)I_3+I_4]$
for $q = u,d$ only.
The contributions to ${\cal{M}}^{\kappa}_{s',s}$
arising from the coefficients for LV at $d=6$
given in cartesian coordinates in the DF are
\begin{align}
{\cal{M}}^{+1}_{\pm\frac12,\pm\frac12}
&=
\pm I_{S2}^q (k^{(6)}_{SV})^z_{qqe\mu}
+I_{S+}^q (k^{(6)}_{SA})^t_{qqe\mu}
\nonumber\\
&
+iI_{V-}^q (k^{(6)}_{VP})^t_{qqe\mu}
\pm iI_{S2}^q (k^{(6)}_{ST})^{tz}_{qqe\mu}
\nonumber\\
&
\pm I_{V2}^q (k^{(6)}_{VV})^{tz}_{qqe\mu}
+I_{V+}^q (k^{(6)}_{VA})^{tt}_{qqe\mu}
\nonumber\\
&
\pm iI_{V2}^q (k^{(6)}_{VT})^{ttz}_{qqe\mu},
\end{align}
\begin{align}
{\cal{M}}^{+1}_{\mp\frac12,\pm\frac12}
&=
I_{S1}^q[(k^{(6)}_{SV})^x_{qqe\mu} \pm i(k^{(6)}_{SV})^y_{qqe\mu}]
\nonumber\\
&
\mp I_{S1}^q[(k^{(6)}_{ST})^{ty}_{qqe\mu} \mp i (k^{(6)}_{ST})^{tx}_{qqe\mu}]
\nonumber\\
&
+I_{V1}^q[(k^{(6)}_{VV})^{tx}_{qqe\mu} \pm i(k^{(6)}_{VV})^{ty}_{qqe\mu}]
\nonumber\\
&
\mp I_{V1}^q[(k^{(6)}_{VT})^{tty}_{qqe\mu} \mp i (k^{(6)}_{VT})^{ttx}_{qqe\mu}],
\end{align}
\begin{align}
{\cal{M}}^{-1}_{\pm\frac12,\pm\frac12}
&=
I_{S+}^q(k^{(6)}_{SV})^t_{qqe\mu}
\pm I_{S2}^q (k^{(6)}_{SA})^z_{qqe\mu}
\nonumber\\
&
+I_{V-}^q (k^{(6)}_{VS})^t_{qqe\mu}
\pm iI_{S2}^q (k^{(6)}_{ST})^{xy}_{qqe\mu}
\nonumber\\
&
+I_{V+}^q  (k^{(6)}_{VV})^{tt}_{qqe\mu}
\pm I_{V2}^q (k^{(6)}_{VA})^{tz}_{qqe\mu}			
\nonumber\\
&
\pm iI_{V2}^q (k^{(6)}_{VT})^{txy}_{qqe\mu},
\end{align}
\begin{align}
{\cal{M}}^{-1}_{\mp\frac12,\pm\frac12}
&=
I_{S1}^q[(k^{(6)}_{SA})^x_{qqe\mu} \pm i(k^{(6)}_{SA})^y_{qqe\mu}]
\nonumber\\
&
+I_{S1}^q[ (k^{(6)}_{ST})^{yz}_{qqe\mu} \mp i (k^{(6)}_{ST})^{xz}_{qqe\mu}]
\nonumber\\
&
+I_{V1}^q[(k^{(6)}_{VA})^{tx}_{qqe\mu} \pm i(k^{(6)}_{VA})^{ty}_{qqe\mu}]
\nonumber\\
&
+I_{V1}^q[ (k^{(6)}_{VT})^{tyz}_{qqe\mu} \mp i (k^{(6)}_{VT})^{txz}_{qqe\mu}].
\end{align}
The conversion rate 
$\omega_{\rm conv} = \overline{|{\cal M}|^2}$ 
corresponding to any given choice of coefficients in the DF 
can be derived directly from these equations.

In parallel with the $d=5$ case,
applying the map~\eqref{map} transforms 
the coefficients for LV in $\omega_{\rm conv}$
from the DF to the SCF,
producing an expression 
involving constant coefficients for LV in the SCF
and time-dependent harmonics of the sidereal frequency $\omega_\oplus$.
Averaging over many sidereal days eliminates
all but the constant term $\overline{\omega}_{\rm conv}$,
which can be used to extract constraints
on components of the coefficients for LV in turn.
The ensuing constraints from SINDRUM~II
and the estimated sensitivities of Mu2e and COMET
are displayed in Table~\ref{resultsd6}.

In the SCF,
the contributions to $\omega_{\rm conv}$ 
proportional to the squared moduli of each coefficient for LV are 
\begin{align}
\omega_{\rm conv}
&\supset
((\zeta_q)_\alpha^{(0)} + (\zeta_q)_\alpha^{(2c)}c_{2\omega_\oplus
T_\oplus}
+ (\zeta_q)_\alpha^{(2s)}s_{2\omega_\oplus T_\oplus})
\nonumber\\
&\qquad\times
( |(k_{SV}^{(6)})^\alpha_{qqe\mu}|^2 +
|(k_{SA}^{(6)})^\alpha_{qqe\mu}|^2 ) 
\nonumber\\
&+ (\zeta^0_q)_T^{(0)}
( |(k_{VS}^{(6)})^T_{qqe\mu}|^2 + |(k_{VP}^{(6)})^T_{qqe\mu}|^2 )
\nonumber\\ &
+\big[((\zeta_q)_{TJ}^{(0)} + (\zeta_q)_{TJ}^{(2c)}c_{2\omega_\oplus
T_\oplus}
+ (\zeta_q)_{TJ}^{(2s)}s_{2\omega_\oplus T_\oplus} )
\nonumber\\
&\qquad\times
|(k_{ST}^{(6)})^{TJ}_{qqe\mu}|^2 
+ \left(TJ\rightarrow JZ\right) \big]
\nonumber\\ &
+\big[(\zeta_q)_{TZ}^{(0)}
|(k_{ST}^{(6)})^{TZ}_{qqe\mu}|^2 + \left(TZ\rightarrow JK\right)\big]
\nonumber\\
&+ ((\zeta^0_q)_{T\alpha}^{(0)}+(\zeta^0_q)_{T\alpha}^{(2c)}c_{2\omega_\oplus
T_\oplus}+(\zeta^0_q)_{T\alpha}^{(2s)}s_{2\omega_\oplus
T_\oplus})
\nonumber\\
&\qquad\times
(|(k_{VV}^{(6)})^{T\alpha}_{qqe\mu}|^2 +
|(k_{VA}^{(6)})^{T\alpha}_{qqe\mu}|^2 )
\nonumber\\
&+ \big[((\zeta^0_q)_{TTJ}^{(0)}+(\zeta^0_q)_{TTJ}^{(2c)}c_{2\omega_\oplus
T_\oplus}+ (\zeta^0_q)_{TTJ}^{(2s)}s_{2\omega_\oplus
T_\oplus})
\nonumber\\
&\qquad\times
|(k_{VT}^{(6)})^{TTJ}_{qqe\mu}|^2 + \left(TTJ\rightarrow
TJZ\right)\big]
\nonumber\\
&+\big[(\zeta^0_q)_{TTZ}^{(0)}|(k_{VT}^{(6)})^{TTZ}_{qqe\mu}|^2 +
\left(TTZ\rightarrow TJK\right)\big],
\label{omconv6}
\end{align}
where a sum over the relevant quark flavors $q$ is understood.
Terms containing products of distinct coefficients for LV are omitted
because they play no role when all coefficient components vanish but one.

\end{document}